\documentstyle[12pt]{article}
\begin{document}
\large
\rightline{UT-707, 1995}
\vfil
\Large
\begin{center}
Analytic Index and Chiral Fermions{%
\renewcommand{\thefootnote}{\fnsymbol{footnote}}
\footnote[2]{Festschrift in honor of Professor H. Banerjee at Saha Institute of
Nuclear Physics (to be published in Indian Jour. of Physics)}
}%
\\
\vfil
Kazuo Fujikawa
\\

Department of Physics, University of Tokyo\\
Bunkyo-ku, Tokyo 113, Japan
\vfil
Abstract

\end{center}

\normalsize

A  recent application of an index relation of the form,
$dim\ ker\ M - dim\ ker\ M^{\dagger} = \nu$,
 to the generation of chiral fermions in a vector-like
gauge theory is reviewed. In this scheme the chiral structure arises
from a mass term with a non-trivial index.The essence of
the generalized Pauli-Villars regularization of chiral gauge theory,which is
based on this mechanism,is also clarified.
\\
\newpage

\section{Introduction}
\par
The notion of index plays an important role in quantum field theory. The best
known example may be the Atiyah-Singer index theorem [1] and chiral anomaly. In
a recent article, Jackiw [2] accounted his encounter with the notion of index
in the study of chiral anomaly and emphasized the importance of various indices
{}.
The Riemann-Roch theorem ,which may be  regarded as a part of the Atiyah-Singer
index theorem ,appears as a ghost number anomaly [3] in two-dimensional quantum
gravity. The Witten index in supersymmetric theory [4] is also related to some
of topological anomalies [5].

 The index associated with a linear operator $M$ is written as
\begin{equation}
dim\ ker\ M - dim\ ker\ M^{\dagger} = \nu
\end{equation}
where $\nu$ stands for an integer and it is called an index. The above form of
index is also called an analytic index. In eq.(1) $dim\ ker\ M$ stands for the
number
of normalizable solutions $u_{n}$ of
\begin{equation}
Mu_{n} = 0
\end{equation}
The index relation (1) is also written as
\begin{equation}
dim\ ker\ M^{\dagger}M - dim\ ker\ MM^{\dagger} = \nu
\end{equation}
The equivalence of these two specifications is seen by noting that $Mu=0$
implies  $M^{\dagger}Mu = 0$. Conversely,$M^{\dagger}Mu = 0$ implies
$(M^{\dagger}Mu, u) = (Mu, Mu) = 0$ and thus $Mu = 0$ if the inner product is
positive definite.

The index is an integer and as such it is expected to be invariant
under a wide class of continuous deformation of parameters characterizing
the operator $M$.

In the present article, I would like to review a recent application
 of
the notion of index to the generation of chiral fermions
via a mass matrix with a non-trivial index. The basic mechanism
of the generalized Pauli-Villars regularization of chiral gauge theories, which
is based on this scheme, is also clarified.

\section{Chiral fermions in a vector-like scheme and a mass matrix with
non-trivial index }
\par
The fundamental fermions appearing in the unified theory of electro-weak
interactions have a chiral structure. At this moment, it is not known how this
chiral structure arises ; it might be that the basic structure of nature has a
chiral structure. On the other hand, in some models of fundamental fermions
such as a vector-like scheme, one envisions the appearance of the chiral
structure as a result of some dynamical effects. Although no definite dynamical
mechanism which realizes this idea   is known, the recent suggestion by
Narayanan and Neuberger[6]on the basis of an analytic index gives an
interesting and suggestive kinematical picture. To be specific, their idea is
to start with a
 vector-like Lagrangian for an $SU(2){ \times} U(1)$ gauge theory, for example,
written in an abbreviated notation
\begin{equation}
{\cal L}_{L}=\overline{\psi}i\gamma^{\mu}D_{\mu}\psi
            - \overline{\psi}_{R}M\psi_{L}
            - \overline{\psi}_{L}M^{\dagger}\psi_{R}
\end{equation}
with
\begin{equation}
\not{\!\! D}=\gamma^{\mu}(\partial_{\mu} - igT^{a}W_{\mu}^{a}
            - i(1/2)g^{\prime}Y_{L}B_{\mu})
\end{equation}
and  $Y_{L}=1/3$ for quarks and $Y_{L}=-1$ for leptons. The field
$\psi$ in (4) is a column vector consisting of an infinite number of
$SU(2)$ doublets, and
the infinite dimensional $nonhermitian$ mass matrix $M$ satisfies the
index condition
\begin{equation}
\dim\ker(M^{\dagger}M) - dim\ker(M M^{\dagger})=3
\end{equation}
and $dim\ker(MM^{\dagger}) = 0$.\\
In the  explicit "diagonalized" expression of $M$
\begin{eqnarray}
M&=&\left(\begin{array}{ccccccc}
          0&0&0&m_{1}&0    &0    &..\\
          0&0&0&0    &m_{2}&0    &..\\
          0&0&0&0    &0    &m_{3}&..\\
          .&.&.&.    &.    &.    &..
          \end{array}\right)\nonumber\\
M^{\dagger}M&=&\left(\begin{array}{cccccc}
          0&&&&&                 \\
           &0&&&0&               \\
           &&0&&&                \\
           &&&m_{1}^{2}&&        \\
           &0&&&m_{2}^{2}&       \\
           &&&&&..
          \end{array}\right)\nonumber\\
M M^{\dagger}&=&\left(\begin{array}{cccccc}
           m_{1}^{2}&&&&&            \\
                    &m_{2}^{2}&&0&&   \\
                    &&m_{3}^{2}&&&   \\
                    &0&&..&&         \\
                    &&&&..&
          \end{array}\right)
\end{eqnarray}
the fermion $\psi$ is written as
\begin{equation}
 \psi_{L}=(1-\gamma_{5})/2\left(
 \begin{array}{c}
  \psi_{1}\\ \psi_{2}\\ \psi_{3}\\ \psi_{4}\\.
 \end{array}
 \right), \ \
 \psi_{R}=(1+\gamma_{5})/2\left(
 \begin{array}{c}
  \psi_{4}\\ \psi_{5}\\ \psi_{6}\\.\\.
 \end{array}
 \right)
\end{equation}
We thus have 3 massless left-handed $SU(2)$ doublets $\psi_{1},\psi_{2},
\psi_{3}$, and an
infinite series of vector-like massive $SU(2)$ doublets $\psi_{4},
\psi_{5},...$ with
masses $m_{1},m_{2},..$ as is seen in\footnote{
One may introduce \underline{constant} complete orthonormal sets
$\{ u_{n} \}$ and $\{ v_{n}\}$ defined by \\
$M^{\dagger}Mu_{n} = 0$ for $n=-2, -1, 0$,\\
$M^{\dagger}Mu_{n} = m_{n}^{2}, M M^{\dagger}v_{n} = m_{n}^{2}v_{n}$
for $n = 1, 2, ...$\\
by assuming the index condition (6). One then has $Mu_{n} = m_{n}v_{n}$
for $m_{n}{\neq} 0$  by choosing the phase of $v_{n}$
and $Mu_{n} = 0$ for $m_{n}=0$ . When one expands\\
$\psi_{L}= \sum_{n=-2}^{\infty} \psi_{n+3}^{L}u_{n},
 \psi_{R}= \sum_{n=1}^{\infty} \psi_{n+3}^{R}v_{n}$\\
one recovers the mass matrix (7) and the relation (9).}

\begin{eqnarray}
{\cal L}_{L}&=&\bar{\psi}_{1}i\not{\!\! D}(\frac{1-\gamma_{5}}{2})\psi_{1}
               +\bar{\psi}_{2}i\not{\!\! D}(\frac{1-\gamma_{5}}{2})\psi_{2}
                \nonumber\\
            & &+\bar{\psi}_{3}i\not{\!\! D}(\frac{1-\gamma_{5}}{2})\psi_{3}
                \nonumber\\
            & &+\bar{\psi}_{4}(i\not{\!\! D} -m_{1})\psi_{4}
               +\bar{\psi}_{5}(i\not{\!\! D} -m_{2})\psi_{5} + ...
\end {eqnarray}

An infinite number of right-handed fermions in a doublet notation are also
introduced by( again in an abbreviated notation)
\begin{equation}
{\cal L}_{R}=\overline{\phi}i\gamma^{\mu}(\partial_{\mu}-i(1/2)g^{\prime}
Y_{R}B_{\mu})\phi - \overline{\phi}_{L}M^{\prime}\phi_{R}
-\overline{\phi}_{R}(M^{\prime})^{\dagger}\phi_{L}
\end{equation}
where
\begin{equation}
Y_{R}=\left(\begin{array}{cc}
            4/3&0\\
            0&-2/3
            \end{array}\right)
\end{equation}
for quarks and
\begin{equation}
Y_{R}=\left(\begin{array}{cc}
            0&0\\
            0&-2
            \end{array}\right)
\end{equation}
for leptons, and the mass matrix $M^{\prime}$   satisfies the index
condition
(6) but in general it may have different mass eigenvalues from
those in(7). After the diagonalization of $M^{\prime}$ with non-zero
eigenvalues $m_{1}^{\prime},m_{2}^{\prime},...$, $\phi$ is
written as
\begin{equation}
 \phi_{L}=(1-\gamma_{5})/2\left(
 \begin{array}{c}
  \phi_{4}\\ \phi_{5}\\ \phi_{6}\\ .\\ .
 \end{array}
 \right), \ \
 \phi_{R}=(1+\gamma_{5})/2\left(
 \begin{array}{c}
  \phi_{1}\\ \phi_{2}\\ \phi_{3}\\ \phi_{4}\\ .
 \end{array}
 \right)
\end{equation}
Here, $\phi_{1}, \phi_{2}$,and  $ \phi_{3}$ are right-handed and massless,
and $\phi_{4}, \phi_{5},....$ have masses $m_{1}^{\prime}, m_{2}^{\prime}$,..
\begin{eqnarray}
{\cal L}_{R}&=&\bar{\phi}_{1}i\not{\!\! D}(\frac{1+\gamma_{5}}{2})\phi_{1}
               +\bar{\phi}_{2}i\not{\!\! D}(\frac{1+\gamma_{5}}{2})\phi_{2}
                \nonumber\\
            & &+\bar{\phi}_{3}i\not{\!\! D}(\frac{1+\gamma_{5}}{2})\phi_{3}
                \nonumber\\
            & &+\bar{\phi}_{4}(i\not{\!\! D} -m_{1}^{\prime})\phi_{4}
               +\bar{\phi}_{5}(i\not{\!\! D} -m_{2}^{\prime})\phi_{5} + ...
\end {eqnarray}
with
\begin{equation}
   \not{\!\! D}= \gamma^{\mu}(\partial_{\mu}-i(1/2)g^{\prime}
                 Y_{R}B_{\mu})
\end{equation}

The present model is vector-like and manifestly anomaly-free
before the  breakdown  of parity (6);after the breakdown of
parity,the model still stays anomaly-free provided that both of $M$ and
$M^{\prime}$ satisfy the index condition (6). In this scheme, the anomaly
 is
caused by the left-right asymmetry, in particular, in the sector of
(infinitely) heavy fermions; in this sense, the parity breaking (6)
may be termed "hard breaking". Unlike  conventional vector-like
models with a finite number of components[7], the present scheme avoids the
appearance of a strongly interacting right-handed sector despite of the
presence of heavy fermions.

The massless fermion sector in the above scheme  reproduces the same
 set of fermions
as in the standard model. However, heavier fermions have distinct
features. For example, the heavier fermion doublets with the smallest
masses are
described by
\begin{eqnarray}
{\cal L}&=&\overline{\psi}_{4}i\gamma^{\mu}(\partial_{\mu}-igW_{\mu}^{a}
           -i(1/2)g^{\prime}Y_{L}B_{\mu})\psi_{4}-m_{1}\overline{\psi}_{4}
           \psi_{4}\nonumber\\
        & &+\overline{\phi}_{4}i\gamma_{\mu}(\partial_{\mu}
           -i(1/2)g^{\prime}Y_{R}B_{\mu})\phi_{4}
           -m_{1}^{\prime}\overline{\phi}_{4}\phi_{4}
\end{eqnarray}
The spectrum of fermions is thus $doubled$ to be vector-like in the
sector  of heavy fermions and ,at the same time, the masses of $\psi$ and
$\phi$ become non-degenerate, i.e., $m_{1}{\neq}m_{1}^{\prime}$.
As a result, the fermion number anomaly[8]
is generated only
by the first 3 generations of light fermions;the violation
of baryon number is not enhanced by the presence of heavier fermions.
The masses  of  heavy doublet components in $\psi$  are degenerate
in the present zeroth order approximation.
If one lets all the masses $m_{1}, m_{2}, ..., m^{\prime}_{1},m^{\prime}_{2},
...$  to $\infty$ in the above model, one recovers the standard model.

Apparently, the present mechanism of generating chiral fermions does not
explain a basic dynamics which is responsible for the chiral
structure. Nevertheless, this kinematical picture is attractive
and might pave a way to a more fundamental understanding of the
chiral structure.

If one assumes that those masses appearing in (9) and (14) are large but
finite, for example, about a few TeV and  heavier, one obtains a generalization
of the conventional vector-like model.
The creation of realistic non-vanishing masses for known light
quarks and leptons , which are
massless in the above scheme, by the Higgs mechanism and the physical
implications of the model are discussed in Ref.[9].

\section{Generalized Pauli-Villars regularization}
\par
The most important feature of the vector-like scheme described in Section 2 is
that all the heavier fremions $decouple$ in the limit of large fermion masses
\begin{eqnarray}
m_{1}, m_{2}, ... & \rightarrow & \infty \nonumber\\
m_{1}^{\prime}, m_{2}^{\prime}, ... & \rightarrow & \infty
\end{eqnarray}
in (9) and (14). In the phenomenological level, this property ensures that
those heavier fermions do not spoil the successful aspects of the
Weinberg-Salam theory.

This decoupling of heavy fermions also implies that those fermions, if suitably
formulated, can be used as regulator fields. In fact, the  recent formulation
of the generalized Pauli-Villars regularization
of chiral gauge theory by Frolov and Slavnov [10] is based on this
property, which in turn led to the vector-like formulation of
Narayanan and Neuberger [6]. To be definite , the chiral theory which we want
to regularize is defined by

\begin{equation}
{\cal L}=\overline{\psi}i\not{\!\!D}\left(\frac{1+\gamma_{5}}{2}\right)\psi
\end{equation}
where
\begin{eqnarray}
\not{\!\!D}&=&\gamma^{\mu}(\partial_{\mu}-igA^{a}_{\mu}(x)T^{a})\nonumber\\
&\equiv&\gamma^{\mu}(\partial_{\mu}-igA_{\mu}(x))
\end{eqnarray}
 In the Euclidean metric we use ,
the Dirac operator $\not{\!\!D}$ is formally hermitian.

The generalized Pauli-Villars regularization of (18) is defined by
\begin{eqnarray}
{\cal L}&=&\overline{\psi}i\not{\!\!D}\psi-\overline{\psi}_{L}M\psi_{R}-
\overline{\psi}_{R}M^{\dag}\psi_{L}\nonumber\\
& &+\overline{\phi}i\not{\!\!D}\phi-\overline{\phi}M'\phi
\end{eqnarray}
where

\begin{eqnarray}
\psi_{R}=\frac{1}{2}(1+\gamma_{5})\psi &,& \psi_{L}=\frac{1}{2}(1-\gamma
_{5})\psi
\end{eqnarray}
and the infinite dimensional mass matrices in (20) are defined by

\begin{eqnarray}
M&=&\left(
\begin{array}{ccccc}
0 & 2 & 0 & 0 & \cdots \\
0 & 0 & 4 & 0 & \cdots \\
0 & 0 & 0 & 6 & \cdots \\
\cdots
\end{array}\right)\Lambda\nonumber\\
M^{\dag}M&=&\left(
\begin{array}{ccccc}
0 &   &   &   &   \\
& 2^{2} & &0 & \\
& &  4^{2} & & \\
&0 &  & 6^{2} & \\
& &  & & \ddots
\end{array}\right)\Lambda^{2}\nonumber\\
MM^{\dag}&=&\left(
\begin{array}{ccccc}
2^{2} &   &      & &   \\
& 4^{2} & &0&  \\
 & & 6^{2} & & \\
 &0 & &  \ddots&\\
 & & & &
\end{array}\right)\Lambda^{2}\nonumber\\
M'&=&\left(
\begin{array}{ccccc}
1 &   &      & &  \\
& 3 & &0 & \\
 & & 5 & & \\
 &0 & &  \ddots&\\
& & & &
\end{array}\right)\Lambda=(M')^{\dag}\nonumber\\
\end{eqnarray}
where $\Lambda$ is a parameter with dimensions of mass.
The mass matrix thus carries a unit index
\begin{equation}
dim\ ker\ M^{\dagger}M - dim\ ker\ MM^{\dagger} = 1
\end{equation}

The fields $\psi$ and $\phi$ in (20) then contain an infinite number
of components , each of which is a conventional 4-component Dirac field;$
\psi(x)$ consists of conventional anti-commuting (Grassmann) fields , and
$\phi(x)$ consists of commuting bosonic Dirac fields.

The Lagrangian (20) is invariant under the gauge transformation
\begin{eqnarray}
\psi(x)&\rightarrow&\psi'(x)=U(x)\psi(x){\equiv}exp[iw^{a}(x)T^{a}]
\psi(x)\nonumber\\
\overline{\psi}(x)&\rightarrow&\overline{\psi}'(x)=\overline{\psi}(x)U(x)
^{\dag}\nonumber\\
\phi(x)&\rightarrow&\phi'(x)=U(x)\phi(x)\nonumber\\
\overline{\phi}(x)&\rightarrow&\overline{\phi}'(x)=
\overline{\phi}(x)U(x)^{\dag}
\nonumber\\
\not{\!\!D}&\rightarrow&\not{\!\!D}'=U(x)\not{\!\!D}U(x)^{\dag}.
\end{eqnarray}
The Noether current associated with the gauge coupling in (20) is
defined by the infinitesimal change of matter variables in (24) with
$\not{\!\!D}$ kept fixed :
\begin{eqnarray}
{\cal L}'&=&\overline{\psi}'i\not{\!\!D}\psi'-\overline{\psi}_{L}'M\psi_{R}'-
\overline{\psi}'_{R}M^{\dag}\psi'_{L}\nonumber\\
&&\ \ \ \ \
+\overline{\phi}'i\not{\!\!D}\phi'-\overline{\phi}'M'\phi'\nonumber\\
&=&-(D_{\mu}w)^{a}J^{{\mu}a}(x)+{\cal L}
\end{eqnarray}
with

\begin{equation}
J^{{\mu}a}(x)=\overline{\psi}(x)T^{a}\gamma^{\mu}\psi(x)+\overline{\phi}
(x)T^{a}\gamma^{\mu}\phi(x).
\end{equation}
Similarly , the U(1) transformation
\begin{eqnarray}
\psi(x)&\rightarrow&e^{i\alpha(x)}\psi(x)\ ,\  \overline{\psi}(x)\rightarrow
\overline{\psi}(x)e^{-i\alpha(x)}\nonumber\\
\phi(x)&\rightarrow&e^{i\alpha(x)}\phi(x)\ ,\  \overline{\phi}(x)\rightarrow
\overline{\phi}(x)e^{-i\alpha(x)}\nonumber\\
\end{eqnarray}
gives rise to the U(1) fermion number current

\begin{equation}
J^{\mu}(x)=\overline{\psi}(x)\gamma^{\mu}\psi(x)+\overline{\phi}(x)
\gamma^{\mu}\phi(x).
\end{equation}
The chiral transformation

\begin{eqnarray}
\psi(x)&\rightarrow&e^{i\alpha(x)\gamma_{5}}\psi(x)\  ,\  \overline{\psi}
\rightarrow\overline{\psi}(x)e^{i\alpha(x)\gamma_{5}}\nonumber\\
\phi(x)&\rightarrow&e^{i\alpha(x)\gamma_{5}}\phi(x)\  ,\  \overline{\phi}
\rightarrow\overline{\phi}(x)e^{i\alpha(x)\gamma_{5}}
\end{eqnarray}
gives the U(1) chiral current

\begin{equation}
J^{\mu}_{5}(x)=\overline{\psi}(x)\gamma^{\mu}\gamma_{5}\psi(x)+
\overline{\phi}(x)\gamma^{\mu}\gamma_{5}\phi(x).
\end{equation}
\par
Considering the variation of action under the transformation (25) and (27)
, one can show that the vector currents (26) and (28) are
$\underline{naively}$ conserved \footnote{The fact that the regularized
currents satisfy anomaly-free relations (31) shows that the regularization
 (20)
is ineffective for the evaluation of possible anomalies in these vector
currents. In particular, this scheme works only for anomaly-free gauge theory.}

\begin{eqnarray}
(D_{\mu}J^{\mu})^{a}(x)&\equiv&\partial_{\mu}J^{{\mu}a}(x)+gf^{abc}A_{\mu}
^{b}(x)J^{{\mu}c}(x)=0,\nonumber\\
\partial_{\mu}J^{\mu}(x)&=&0
\end{eqnarray}
whereas the chiral current (30) satisfies the $\underline{naive}$
identity
\begin{equation}
\partial_{\mu}J^{\mu}_{5}(x)=2i\overline{\psi}_{L}M\psi_{R}-2i\overline{\psi}
_{R}M^{\dag}\psi_{L}+2i\overline{\phi}M'\gamma_{5}\phi.
\end{equation}

\par
The quantum theory of (20) may be defined by the path integral  , for example ,

\begin{equation}
<\overline{\psi}(x)T^{a}\gamma^{\mu}\psi(x)>={\int}d\mu\overline{\psi}
(x)T^{a}\gamma^{\mu}\psi(x)exp[{\int}{\cal L}d^{4}x].
\end{equation}
The path integral over the bosonic variables $\phi$ and $\overline{\phi}$
for the Dirac operator in Euclidean theory needs to be defined via a
suitable rotation in the functional space.

\bigskip

\begin{flushleft}
\underline{\bf{Definition of currents in terms of propagators}}
\end{flushleft}
\par
We now define the currents in terms of propagators to clarify the basic
mechanism of generalized Pauli-Villars regularization [11].
The basic idea of this approach is explained for the $\underline{un-
regularized}$ theory in (18) as follows : We start with the current
associated with the gauge coupling

\begin{eqnarray}
\lefteqn{<\overline{\psi}(x)T^{a}\gamma^{\mu}(\frac{1+\gamma_{5}}{2})
\psi(x)>}\nonumber\\
&=&\lim_{y{\rightarrow}x}<T^{*}\overline{\psi}(y)T^{a}\gamma^{\mu}
(\frac{1+\gamma_{5}}{2})\psi(x)>\nonumber\\
&=&-\lim_{y{\rightarrow}x}<T^{*}(T^{a})_{bc}\gamma^{\mu}_{{\alpha}\delta}
(\frac{1+\gamma_{5}}{2})_{\delta\beta}\psi_{{\beta}c}(x)\overline{\psi}
_{{\alpha}b}(y)>\nonumber\\
&=&\lim_{y{\rightarrow}x}Tr[T^{a}\gamma^{\mu}(\frac{1+\gamma_{5}}{2})
\frac{1}{i\not{\!\!D}}\delta(x-y)]
\end{eqnarray}
where we used the anti-commuting property of $\psi$ and the expression of
 the propagator

\begin{equation}
<T^{*}\psi(x)\overline{\psi}(y)>=(\frac{1+\gamma_{5}}{2})\frac{(-1)}
{i\not{\!\!D}_{x}}\delta(x-y)
\end{equation}
The trace in (34) runs over the Dirac and Yang-Mills indices.
We now notice the expansion

\begin{eqnarray}
\frac{1}{i\not{\!\!D}}&=&\frac{1}{i\not{\!\partial}+g\not{\!\!A}}\nonumber\\
&=&\frac{1}{i\not{\!\partial}}+\frac{1}{i\not{\!\partial}}(-g
\not{\!\!A})\frac{1}{i\not{\!\partial}}\nonumber\\
&&+\frac{1}{i\not{\!\partial}}(-g
\not{\!\!A})\frac{1}{i\not{\!\partial}}(-g\not{\!\!A})
\frac{1}{i\not{\!\partial}}+\cdots
\end{eqnarray}
When one inserts (36) into (34) and retains only the terms linear in
$A^{b}_{\nu}(x)$ , one obtains

\begin{eqnarray}
\lefteqn{\lim_{y{\rightarrow}x}Tr[T^{a}\gamma^{\mu}(\frac{1+\gamma_{5}}{2})
\frac{(-1)}{i\not{\!\partial}}\gamma^{\nu}T^{b}gA^{b}_{\nu}(x)
\frac{1}{i\not{\!\partial}}\delta(x-y)]}\nonumber\\
&=&\lim_{y{\rightarrow}x}{\int}d^{4}zTr[T^{a}\gamma^{\mu}(\frac{1+
\gamma_{5}}{2})\frac{(-1)}{i\not{\!\partial}}\nonumber\\
&& \ \ \ {\times}\delta(x-z)T^{b}\gamma^{\nu}
\frac{1}{i\not{\!\partial}}\delta(x-y)]gA^{b}_{\nu}(z)
\end{eqnarray}
where the derivative $\partial_{\mu}$ acts on $\underline{all}$ the x-
variables standing on the right of it in (37).
If one takes the variational derivative of (37) with respect to $gA
^{b}_{\nu}(z)$ , one obtains

\begin{eqnarray}
\lefteqn{\lim_{y{\rightarrow}x}Tr[T^{a}\gamma^{\mu}(\frac{1+\gamma_{5}}{2})
\frac{(-1)}{i\not{\!\partial}}\delta(x-z)\gamma^{\nu}T^{b}
\frac{1}{i\not{\!\partial}}\delta(x-y)]}\nonumber\\
&=&\lim_{y{\rightarrow}x}\int\frac{d^{4}q}{(2\pi)^{4}}\frac{d^{4}k}
{(2\pi)^{4}}Tr[T^{a}
\gamma^{\mu}(\frac{1+\gamma_{5}}{2})\frac{(-1)}{\not{\!k}+\not{\!q}}
T^{b}\gamma^{\nu}\frac{1}{\not{\!k}}]e^{-iq(x-z)}e^{-ik(x-y)}\nonumber\\
&=&\int\frac{d^{4}q}{(2\pi)^{4}}e^{-iq(x-z)}(-1)
\int\frac{d^{4}k}{(2\pi)^{4}}Tr[T^{a}
\gamma^{\mu}(\frac{1+\gamma_{5}}{2})\frac{1}{\not{\!k}+\not{\!q}}
T^{b}\gamma^{\nu}(\frac{1+\gamma_{5}}{2})\frac{1}{\not{\!k}}]\nonumber\\
&\equiv&\int\frac{d^{4}q}{(2\pi)^{4}}e^{-iq(x-z)}\Pi^{ab}_{\mu\nu}(q)
\end{eqnarray}
where we used the representations of $\delta$-function

\begin{eqnarray}
\delta(x-z)&=&\int\frac{d^{4}q}{(2\pi)^{4}}e^{-iq(x-z)}\nonumber\\
\delta(x-y)&=&\int\frac{d^{4}k}{(2\pi)^{4}}e^{-ik(x-y)}.
\end{eqnarray}
\par
The last expression in (38) stands for the vacuum polarization tensor.
Namely , one can generate the multiple correlation functions of currents
$\overline{\psi}T^{a}\gamma^{\mu}(\frac{1+\gamma_{5}}{2})\psi$ in the
perturbative sense by taking the variational derivative of (34) with
respect to gauge fields $A_{\mu}^{a}$ .
This idea also works for the non-gauge currents (28) and (30).
We emphasize that we always take the limit $y=x$ first before the
explicit calculation , and thus (34) \underline{differs}
 from the point-splitting
definition of currents.
\par
We now generalize the above definition of currents for the theory
defined by (20).
For this purpose , we rewrite (20) as
\begin{equation}
{\cal L}=\overline{\psi}i{\cal D}\psi+\overline{\phi}i{\cal D}'\phi
\end{equation}
with
\begin{eqnarray}
\cal D&\equiv&\not{\!\!D}+iM(\frac{1+\gamma_{5}}{2})+iM^{\dag}
(\frac{1-\gamma_{5}}{2})\nonumber\\
\cal D'&\equiv&\not{\!\!D}+iM'.
\end{eqnarray}
The gauge current (26) is then defined by

\begin{eqnarray}
J^{{\mu}a}(x)&=&\lim_{y{\rightarrow}x}\{<T^{*}\overline{\psi}(y)T^{a}
\gamma^{\mu}\psi(x)>+<T^{*}\overline{\phi}(y)T^{a}
\gamma^{\mu}\phi(x)>\}\nonumber\\
&=&\lim_{y{\rightarrow}x}\{-<T^{*}T^{a}\gamma^{\mu}\psi(x)\overline{\psi}
(y)>+<T^{*}T^{a}\gamma^{\mu}\phi(x)\overline{\phi}(y)>\}\nonumber\\
&=&\lim_{y{\rightarrow}x}Tr[T^{a}\gamma^{\mu}(\frac{1}{i\cal D}-\frac{1}{i\cal
D'})
\delta(x-y)]
\end{eqnarray}
where trace includes the sum over the infinite number of field components
 in addition to Dirac and Yang-Mills indices.
 The anti-commuting property of $\psi(x)$ and the commuting property of $
\phi(x)$ are used in (42).
\par
We next notice the relations

\begin{eqnarray}
\frac{1}{\cal D}&=&\frac{1}{\cal D^{\dag}\cal D}\cal D^{\dag}\nonumber\\
&=&\frac{1}{\not{\!\!D}^{2}+\frac{1}{2}M^{\dag}M(1+\gamma_{5})+\frac{1}{2}
MM^{\dag}(1-\gamma_{5})}\cal D^{\dag}\nonumber\\
&=&[(\frac{1+\gamma_{5}}{2})\frac{1}{\not{\!\!D}^{2}+M^{\dag}M}+(\frac{
1-\gamma_{5}}{2})\frac{1}{\not{\!\!D}^{2}+MM^{\dag}}]\nonumber\\
&&\ \ \ \ \ \times[\not{\!\!D}-iM^{\dag}(\frac{
1+\gamma_{5}}{2})-iM(\frac{1-\gamma_{5}}{2})]\nonumber\\
\frac{1}{\cal D'}&=&\frac{1}{(\cal D')^{\dag}\cal D'}(\cal
D')^{\dag}\nonumber\\
&=&\frac{1}{\not{\!\!D}^{2}+(M')^{2}}(\not{\!\!D}-iM').
\end{eqnarray}
We thus rewrite (42) as

\begin{eqnarray}
\lefteqn{Tr\left[-iT^{a}\gamma^{\mu}(\frac{1}{\cal D}-\frac{1}{\cal D'})
\delta(x-y)\right]}
\nonumber\\
&=&Tr\left\{-iT^{a}\gamma^{\mu}\left[(\frac{1+\gamma_{5}}{2})\sum^{\infty}
_{n=0}
\frac{1}{\not{\!\!D}^{2}+(2n\Lambda)^{2}}\right.\right.\nonumber\\
&&\ \ \ \ \ \ \ \ +(\frac{1-\gamma_{5}}{2})
\sum^{\infty}_{n=1}\frac{1}{\not{\!\!D}^{2}+(2n\Lambda)^{2}}\nonumber\\
& &\ \ \ \ \ \ \ \  \left.\left.-\sum^{\infty}_{n=0}\frac{1}{\not{\!\!D}
^{2}+[(2n+1)\Lambda]^{2}}\right]
\not{\!\!D}\delta(x-y)\right\}\nonumber\\
&=&\frac{1}{2}Tr\left[-iT^{a}\gamma^{\mu}\sum^{\infty}_{n=-\infty}
\frac{(-1)^{n}\not{\!\!D}^{2}}{\not{\!\!D}^{2}+(n\Lambda)^{2}}
\frac{1}{\not{\!\!D}}\delta(x-y)\right]\nonumber\\
& &\ \ \ \ \ \ \ \ +\frac{1}{2}Tr\left[-iT^{a}\gamma^{\mu}\gamma_{5}
\frac{1}{\not{\!\!D}}
\delta(x-y)\right]\nonumber\\
&=&\frac{1}{2}Tr\left[T^{a}\gamma^{\mu}f(\not{\!\!D}^{2}/\Lambda^{2})
\frac{1}{i\not{\!\!D}}\delta(x-y)\right]\nonumber\\
& &\ \ \ \ \ \ \ \ +\frac{1}{2}Tr\left[T^{a}\gamma^{\mu}
\gamma_{5}\frac{1}{i\not{\!\!D}}
\delta(x-y)\right]
\end{eqnarray}
where we explicitly evaluated the trace over the infinite number of components
and used the fact that the trace over an odd number of $\gamma$-matrices
 vanishes.
We also defined $f(x^{2})$ by

\begin{eqnarray}
f(x^{2})&\equiv&\sum^{\infty}_{n=-\infty}\frac{(-1)^{n}x^{2}}{x^{2}+
(n\Lambda)^{2}}\nonumber\\
&=&\frac{(\pi{x}/\Lambda)}{sinh({\pi}x/\Lambda)}.
\end{eqnarray}
This last expression of (45) as a sum of infinite number of terms is
given in Ref.[10].
The regulator $f(x^{2})$, which rapidly approaches 0 at $x^{2}=\infty$,
satisfies

\begin{eqnarray}
f(0)&=&1\nonumber\\
x^{2}f'(x^{2})&=&0 \ for\   x\rightarrow{0}\nonumber\\
f(+\infty)&=&f'(+\infty)=f''(+\infty)=\cdots=0\nonumber\\
x^{2}f'(x^{2})&\rightarrow&0 \ \  for \ \  x\rightarrow\infty.
\end{eqnarray}
\par
The essence of the generalized Pauli-Villars regularization (20) is thus
 summarized in terms of regularized currents as follows:

\begin{eqnarray}
\lefteqn{<\overline{\psi}(x)T^{a}\gamma^{\mu}(\frac{1+\gamma_{5}}{2})\psi(x)>
_{PV}}\nonumber\\
&=&\lim_{y{\rightarrow}x}\left\{\frac{1}{2}Tr\left[T^{a}
\gamma^{\mu}f(\not{\!\!D}^{2}
/\Lambda^{2})\frac{1}{i\not{\!\!D}}\delta(x-y)\right]\right.\nonumber\\
& &\ \
\left.+\frac{1}{2}Tr\left[T^{a}\gamma^{\mu}\gamma_{5}
\frac{1}{i\not{\!\!D}}\delta
(x-y)\right]\right\}\\
\lefteqn{<\overline{\psi}(x)\gamma^{\mu}(\frac{1+\gamma_{5}}{2})\psi(x)>_{PV}}
\nonumber\\
&=&\lim_{y{\rightarrow}x}\left\{\frac{1}{2}Tr\left[\gamma^{\mu}
f(\not{\!\!D}^{2}/\Lambda^{2})
\frac{1}{i\not{\!\!D}}\delta(x-y)\right]\right.\nonumber\\
& &\ \ \left.+\frac{1}{2}Tr\left[\gamma^{\mu}\gamma_{5}\frac{1}{i\not{\!\!D}}
\delta(x-y)\right]\right\}\\
\lefteqn{<\overline{\psi}(x)\gamma^{\mu}\gamma_{5}(\frac{1+\gamma_{5}}{2})
\psi(x)>_{PV}}\nonumber\\
&=&\lim_{y{\rightarrow}x}\left\{\frac{1}{2}Tr\left[
\gamma^{\mu}\gamma_{5}f(\not{\!\!D}^{2}/\Lambda^{2})
\frac{1}{i\not{\!\!D}}\delta(x-y)\right]\right.\nonumber\\
& &\ \ \left.+\frac{1}{2}Tr\left[\gamma^{\mu}\frac{1}
{i\not{\!\!D}}\delta(x-y)\right]\right\}.
\end{eqnarray}
In the left-hand sides of (47)$\sim$(49), the currents are defined in terms of
the original fields appearing in (18).
The vector $U(1)$ and  axial-vector  currents written in terms of the original
fields in (18) are identical , but the regularized versions ,i.e.,
(48) and (49) are different. In particular , the vector $U(1)$
 current,i.e.,(48) is not completely regularized.
See also Refs.[6] and [12]. This reflects the different form of
\underline{naive} identities in (31) and (32) ; if all the currents
are well regularized , the \underline{naive} form of identities would
also coincide. We emphasize that all the one-loop diagrams are generated
 from the (partially) regularized currents in (47) $\sim$ (49) ; in other words
,(47) $\sim$ (49)
retain all the information of the generalized Pauli-Villars
regularization (20).

\section{Generalized Pauli-Villars regularization and anomalies}
\par
As is seen in (47), the possible anomalous term of the gauge current
which contains $\gamma_{5}$ is not regularized. The generalized Pauli-Villars
regularization of chiral gauge theory thus works only for the theories which
contain no gauge anomaly[10]. In an anomaly-free
gauge theory such as the Weinberg-Salam theory, the U(1) fermion
number anomaly is physical and interesting. In the generalized Pauli-Villars
regularization in (20), the possible anomalous term of the
U(1) current
\begin{eqnarray}
<\overline{\psi}(x)\gamma^{\mu}(\frac{1+\gamma_{5}}{2})\psi(x)>_{PV}
\end{eqnarray}
is not regularized, since the term which contains $\gamma_{5}$ is not
regularized in (48). On the other hand, the ``axial-current''
\begin{equation}
<\overline{\psi}(x)\gamma^{\mu}\gamma_{5}(\frac{1+\gamma_{5}}{2})
\psi(x)>_{PV}
\end{equation}
which is identical to (50) in the un-regularized theory, is in fact
different in the generalized Pauli-Villars regularization in (49) and the
possible anomalous term containing $\gamma_{5}$ is regularized as
\begin{eqnarray}
\lefteqn{<\overline{\psi}(x)\gamma^{\mu}\gamma_{5}(\frac{1+\gamma_{5}}{2})
\psi(x)>_{PV}}\nonumber\\
&=&\lim_{y{\rightarrow}x}\left\{\frac{1}{2}Tr\left[
\gamma^{\mu}\gamma_{5}f(\not{\!\!D}^{2}/\Lambda^{2})
\frac{1}{i\not{\!\!D}}\delta(x-y)\right]\right.\nonumber\\
& &\ \ \left.+\frac{1}{2}Tr\left[\gamma^{\mu}\frac{1}
{i\not{\!\!D}}\delta(x-y)\right]\right\}\nonumber\\
&\rightarrow&\sum_{n}\phi_{n}(x)^{\dag}\left[\gamma^{\mu}(\frac{\gamma_{5}}{2})f(
\lambda_{n}^{2}/\Lambda^{2})\frac{1}{i\lambda_{n}}\right]\phi_{n}(x)
\end{eqnarray}
where we used the complete set defined by
\begin{eqnarray}
\not{\!\!D}\phi_{n}(x)&\equiv&\lambda_{n}\phi_{n}(x)\nonumber\\
\int\phi_{m}(x)^{\dag}\phi_{n}(x)d^{4}x&=&\delta_{m,n}\nonumber\\
\delta_{\alpha\beta}\delta(x-y)&\rightarrow&\sum_{n}\phi_{n}(x)_{\alpha}
\phi_{n}(y)^{\dag}_{\beta}
\end{eqnarray}
with $\alpha$ and $\beta$ including Dirac and Yang-Mills indices.
One can thus evaluate the fermion number anomaly by using the last
expression of the
``axial-current'' (52) as
\begin{eqnarray}
\lefteqn{\partial_{\mu}<\overline{\psi}(x)\gamma^{\mu}\gamma_{5}(\frac{1+\gamma_{5}}{2})
\psi(x)>_{PV}}\nonumber\\
&=&\sum_{n}\left[-(\not{\!\!D}\phi_{n}(x))^{\dag}(\frac{\gamma_{5}}{2})
f(\lambda_{n}^{2}/\Lambda^{2})\frac{1}{i\lambda_{n}}\phi_{n}(x)\right.
\nonumber\\
& &\ \ \ \ \ \left.+\phi_{n}(x)^{\dag}(\frac{-\gamma_{5}}{2})
f(\lambda_{n}^{2}/\Lambda^{2})\frac{1}{i\lambda_{n}}(\not{\!\!D}\phi_{n}
(x))\right]\nonumber\\
&=&i\sum_{n}\phi_{n}(x)^{\dag}\gamma_{5}f(\lambda^{2}_{n}/\Lambda^{2})\phi
_{n}(x)\nonumber\\
&=&iTr\int\frac{d^{4}k}{(2\pi)^{4}}e^{-ikx}\gamma_{5}f(\not{\!\!D}^{2}/
\Lambda^{2})e^{ikx}\nonumber\\
&=&(\frac{ig^{2}}{32\pi^{2}})Tr\epsilon^{\mu\nu\alpha\beta}F_{\mu\nu}
F_{\alpha\beta} \ \  for\ \   \Lambda\rightarrow\infty
\end{eqnarray}
where we used the relation $(\gamma_{\mu})^{\dagger}= -\gamma_{\mu}$
in the present Euclidean metric. We also followed the calculational scheme in
Ref.[13] in the last step of (54).

We thus recover the conventional covariant form of anomaly for the
fermion number current [8]. This calculation of the fermion number anomaly in
the generalized Pauli-Villars regularization was first performed by Aoki and
Kikukawa [12] on the basis of Feynman diagrams.

\section{Conclusion}
The chiral structure is the most fundamental property of elementary
fermions in modern unified gauge theory. The chiral anomaly, which is related
to the chiral structure, is a subtle but profound phenomenon in field theory.
It is interesting that the generalized Pauli-Villars
regularization [10] successfully  regularizes the Weinberg-Salam theory,
although it requires an infinite number of regulator fields.

The notion of chiral anomaly is also closely related to the so-called U(1)
problem and strong CP problem. In this connection, H. Banerjee and his
collaborators
continuously clarified the most fundamental aspects of the problem and
suggested a careful reassessment of the path integration quantization of modern
gauge theory itself [14].

On the occasion of the 60th birthday of Prof. H. Banerjee, I wish him many
happy returns.

\end{document}